%% file: colm2024_conference.tex
\documentclass{article} 
\usepackage{colm2024_conference}

\usepackage{microtype}
\usepackage{hyperref}
\usepackage{url}
\usepackage{booktabs}
\usepackage{multirow}
\usepackage{hyperref}
\usepackage{amsmath}
\usepackage{amssymb}
\usepackage{mathtools}
\usepackage{amsthm}
\usepackage{subcaption}
\usepackage[capitalize,noabbrev]{cleveref}
\usepackage{floatrow}
\newfloatcommand{capbtabbox}{table}[][\FBwidth]

\theoremstyle{plain}

\theoremstyle{definition}

\theoremstyle{remark}

\usepackage{wrapfig}

\title{MuPT: A Generative Symbolic Music Pretrained Transformer}

\author{Xingwei Qu$^1$\,$^3$\,$^4$\footnotemark[1]\;,
    Yuelin Bai$^5$\footnotemark[1]\;,
    Yinghao Ma$^1$\,$^7$\footnotemark[1]\;,\\
    \textbf{Ziya Zhou}$^3$,
    \textbf{Ka Man Lo}$^3$,
    \textbf{Jiaheng Liu}$^1$,
    \textbf{Ruibin Yuan}$^1$\,$^3$,
    \textbf{Lejun Min}$^8$,
    \textbf{Xueling Liu}$^1$,\\
    \textbf{Tianyu Zhang}$^9$,
    \textbf{Xinrun Du}$^1$,
    \textbf{Shuyue Guo}$^1$,
    \textbf{Yiming Liang}$^{10}$,
    \textbf{Yizhi Li}$^1$\,$^4$,
    \textbf{Shangda Wu}$^{11}$,\\
    \textbf{Junting Zhou}$^{12}$,
    \textbf{Tianyu Zheng}$^1$,
    \textbf{Ziyang Ma}$^{13}$,
    \textbf{Fengze Han}$^1$,
    \textbf{Wei Xue}$^3$,
    \textbf{Gus Xia}$^8$,\\
    \textbf{Emmanouil Benetos}$^7$,
    \textbf{Xiang Yue}$^1$,
    \textbf{Chenghua Lin}$^4$,
    \textbf{Xu Tan}$^{14}$,
    \textbf{Stephen W. Huang}$^{15}$\\
    \textbf{Jie Fu}$^3$\footnotemark[2]\;,
    \textbf{Ge Zhang}$^1$\,$^2$\,$^6$\footnotemark[1] \, \footnotemark[2] \\[2mm]
    $^1$M-A-P, $^2$University of Waterloo, $^3$HKUST, $^4$University of Manchester, \\$^5$Shenzhen Institute of Advanced Technology, CAS, $^6$Vector Institue, $^7$QMUL, $^8$MBZUAI,\\ $^9$MILA, $^{10}$Institute of Automation, CAS, $^{11}$Central Conservatory of Music, \\$^{12}$PKU, $^{13}$SJTU, $^{14}$MSRA, $^{15}$harmony.ai
\\[2mm]
}

\colmfinalcopy 
\begin{document}

\maketitle
\let\thefootnote\relax\footnotetext{*$\ $ Equal Technical Contributions.}
\let\thefootnote\relax\footnotetext{\textsuperscript{\dag} $\ $Corresponding Authors.}

\vspace{-7ex}
\begin{center}
     \url{https://map-mupt.github.io/}
\end{center}
\vspace{0ex}

\begin{abstract}

In this paper, we explore the application of Large Language Models (LLMs) to the pre-training of music. While the prevalent use of MIDI in music modeling is well-established, our findings suggest that LLMs are inherently more compatible with ABC Notation, which aligns more closely with their design and strengths, thereby enhancing the model's performance in musical composition.
To address the challenges associated with misaligned measures from different tracks during generation, we propose the development of a \underline{S}ynchronized \underline{M}ulti-\underline{T}rack ABC Notation (\textbf{SMT-ABC Notation}), which aims to preserve coherence across multiple musical tracks. 
Our contributions include a series of models capable of handling up to 8192 tokens, covering 90\% of the symbolic music data in our training set. Furthermore, we explore the implications of the \underline{S}ymbolic \underline{M}usic \underline{S}caling Law (\textbf{SMS Law}) on model performance. The results indicate a promising direction for future research in music generation, offering extensive resources for community-led research through our open-source contributions.

\end{abstract}

\section{Introduction}

Large Language Models (LLMs) have experienced remarkable advancements, leading to their broad application across numerous domains. As these models extend into multimodal areas, such as visual and auditory fields, their capability to represent and model complex information, including images \citep{liu2023visual} and speech \citep{baevski2020wav2vec} becomes increasingly critical. However, this expansion also highlights significant challenges that must be addressed. Specifically, the development of effective tokenizers for images and videos, as well as advanced codecs for the audio domain. 

In the domain of music, Large Language Models encounter inherent challenges that hinder their effective utilization. Despite achieving state-of-the-art musical performance, as demonstrated by MuseNet~\citep{MuseNet}, these models often struggle to capture the structural symmetry essential to aesthetically pleasing music. This issue stems from the use of Musical Instrument Digital Interface (MIDI), which, while effective, poses significant challenges in terms of music's readability and structural representation. 

To tackle this issue, the integration of ABC notation offers a novel approach to overcoming the limitations of MIDI formats. \citet{yuan2024chatmusician} advocate for this method, highlighting ABC notation's readability and structural coherence. Their methodology involves fine-tuning the LLAMA2 model, leveraging instruction tuning to enhance the model's musical output capabilities~\citep{touvron2023llama2, touvron2023llama}. The research overlooks critical tokenization considerations within musical compositions. 

In this paper, we aim to propose a training standard with transformer decoder-only architecture for symbolic music generation tasks, which is suitable for single / multi-track music generation. We observe that mismatches between measures can occur by employing the traditional 'next-token-prediction' paradigm for symbolic data training. This issue arises because ABC notations are generally notated track by track, completing one track before moving on to the next. To address this challenge, we propose SMT-ABC notation to facilitate the model's learning of how each measure is expressed across various tracks. 

Furthermore, we observe that the ABC Notation model benefits from additional epochs in the training phase. This suggests that repeated data positively impacts the model's performance. To understand this phenomenon, we introduced the SMS Law for repetitive training with symbolic music data. This law explores how scaling up the training data affects the performance of symbolic music generation models, particularly in terms of validation loss. This investigation aims to provide clear insights into the relationship between data repetition and model efficacy, offering guidance for optimizing model training strategies.


In conclusion, our contributions are highlighted as follows:
\begin{itemize}
    \item We develop a Long-range Symbolic Music LLM that introduced a foundation model trained on musical notes in ABC notation, with an extended sequence length of 8192 tokens, catering to over 90\% of symbolic musical scores we collected.
    \item We propose SMT-ABC notation to represent notes, significantly improving the structural integrity and quality of the generated music by maintaining consistent measures within each track.
    \item We explore the SMS Law insights for ABC notation. We demonstrate that comprehensive song modeling yields superior performance with a positive correlation between model size and metric improvement. We also reveal unique training epoch dynamics in music repetition and performance enhancement.
    \item We will release a suite of state-of-the-art long-range foundation models in the music domain, articulated in ABC notation, along with intermediate training checkpoints to foster community research and innovation in symbolic music modeling.
\end{itemize}

\section{Related work}

\subsection{Music Pre-training}

Audio pre-training through the self-supervised learning paradigm has made great progress in speech~\citep{baevski2020wav2vec,hsu2021hubert,baevski2022data2vec, ma2022mt4ssl, yang2023fast, lin2023melhubert}, general-purpose audio~\citep{huang2022masked, baade2022mae, chen2023beats, chen2024eat}, as well as music~\citep{zhu2021musicbert, dong2023multitrack, thickstun2023anticipatory, ma2023effectiveness, li2023mert}. 
Two types of self-supervised music pre-training have been explored: non-autoregressive discriminative models and autoregressive generative models. 
Non-autoregressive discriminative music pre-training performs mask language modelling (MLM) by constructing a pretext task. 
This kind of training makes models easier to adapt to downstream understanding tasks, such as music tagging, instrument classification, and beat tracking.
Autoregressive generative music pre-training models employ a GPT-style framework to generate music, either in codec~\citep{copet2024simple} form or in symbolic form~\citep{thickstun2023anticipatory, dong2023multitrack}. 
Previous symbolic music generation models utilize MIDI to model the sequence input and output, showing the ability to generate music given conditions, or unconditional generation. 
Existing models are limited by not generating long enough music~\citep{thickstun2023anticipatory} and limited musicality~\citep{dong2023multitrack}. 
Therefore, long-range symbolic music generation models with data scaling and model scaling need to be explored urgently. 

\subsection{Data Representation for Symbolic Music}

Symbolic music representation formats such as MIDI, Humdrum, and ABC notation offer distinct approaches for representing musical information, each with unique advantages and applicability to computational music representation. MIDI, which excels in capturing musical notes and performance, is a popular choice in the music industry and research community\citep{huang2020pop,huang2018music,lu2023musecoco}. However, the complexity and length of MIDI sequences often challenge music models, which limit the preservation of a composition’s full continuity. 
In contrast, ABC notation stands out for its textual simplicity and compactness, making it particularly suited for Natural Language Processing (NLP) techniques. It can be efficiently processed and analyzed using sequence modeling and pattern recognition algorithms similar to those used in language translation and text generation, enabling automated music generation and retrieval.

ABC notation's simplicity and broad applicability have prompted research into enhancing music retrieval and generation through deep learning and NLP. In early research, LSTM networks showed promise by producing music similar to traditional and folk styles \citep{DBLP:journals/corr/SturmSBK16}, using ABC notation for automated composition. Following this, TunesFormer \citep{DBLP}, a tool based on the Transformer designed for Irish tunes encoded in ABC notation, utilized techniques like bar patching and introduced control codes to craft melodies that meet specific musical forms. abcMLM \citep{casini2023generating}, a masked language model, further demonstrated how structured ABC notation can be used to create folk-like tunes, highlighting the adaptability of NLP methods and the benefits of using non-autoregressive models in music. Recent studies have started to utilize pre-trained NLP models for converting text to music \citep{wu2023exploring}, showing how these resources can improve the creation of symbolic music. CLaMP \citep{WuY0S23} introduced a unique method for learning music and text jointly, using a large collection of music-text pairs to better search for and categorize music automatically. Techniques like text dropout and bar patching are examples of how NLP and music encoding are becoming more integrated. In a significant breakthrough, ChatMusician \citep{yuan2024chatmusician} introduced a new approach to incorporating music as a second language for Large Language Models (LLMs), utilizing ABC notation to seamlessly blend music and text, thereby enabling internal music creation and analysis without relying on external multimodal frameworks.

\subsection{Scaling Law}
A wide range of research underscores a significant pattern in language model performance, indicating a power-law relationship between model performance and the increases in both the number of parameters and the size of the training data \citep{kaplan2020scaling, hoffmann2022training, ghorbani2021scaling}. Scaling law plays a pivotal role in advancing large language models (LLMs), offering a framework to predict the optimal configurations for larger models based on the training logs of their smaller counterparts \citep{gao2022scaling}.

Further exploration into scaling laws for autoregressive generative modeling by \citet{henighan2020scaling} broadens the applicability of these laws to include not just textual, but also visual and multimodal tasks, as supported by studies in \citet{ghorbani2021scaling, hernandez2021scaling, gao2022scaling}. Such insights are invaluable for developing music generation models, which often blend multiple modalities such as audio, lyrics, and visual elements like album covers or artist photos. This demonstrates a consistent trajectory of performance enhancement concurrent with resource scaling.

The research by \citet{muennighoff2024scaling}, which involves the repetition of the entire pre-training dataset across multiple epochs, presents promising results yet raises questions regarding its effectiveness for musical data. This uncertainty prompts a need for further research into the impact of data repetition strategy by achieving improved outcomes for models engaged in music-related tasks.


\section{Method}
\subsection{Model Architecture}\label{sec:model arch}

MuPT utilizes a standard Transformer model architecture \citep{vaswani2023attention} in a decoder-only setup. Models are trained on a context
length of 8192 tokens. We list our MuPT model parameter in Table \ref{tab:model_parameters} and utilize several improvements proposed after the original transformer paper. Below, we list the included improvements:
\begin{itemize}
    \item \textbf{SwiGLU Activation:} The SwiGLU activation mechanism, represented as $(\text{Swish}(xW) \cdot xV)$, is utilized for the MLP (Multi-Layer Perceptron) intermediate activations. This approach significantly surpasses traditional activation functions such as ReLU, GeLU, and Swish in performance \citep{shazeer2020glu}.
    

    \item \textbf{RMSNorm} Each transformer sub-layer, including the attention and feedforward layers, is normalized using RMSNorm as proposed by \citet{zhang2019root}
    
    \item \textbf{RoPE Embeddings:} In contrast to positional encoding (PE) strategy, we use the Rotary Positional Encoding (RoPE) technique, as developed by \citet{su2023roformer}, aimed at enhancing long-context modeling.

\end{itemize}

\subsection{SMT-ABC Notation}

ABC notation is a widely adopted system for notating music using plain text, and it offers unique advantages when used in conjunction with deep learning models. 
Its well-structured text format ensures easy preprocessing, efficient data transmission, and scalability of datasets. 
The diverse collection of tunes and compositions in ABC notation facilitates learning various musical structures and styles.
Moreover, ABC notation allows models to generate human-readable outputs, leading to immediate feedback and iterative refinement.
These attributes significantly enhance both the efficiency and quality of the training process. 

\begin{wrapfigure}[]{r}{.36\textwidth}
    \begin{center}
        \fbox{\includegraphics[width=.94\textwidth]{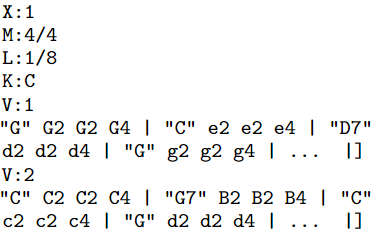}}
    \end{center}
    \caption{Example of a multi-track tune of ABC notation.}
    \label{fig:abc_sample}
\end{wrapfigure}

An ABC file is composed of headers followed by the music notation. 
The former contain metadata regarding the tune, such as its composer and tempo, while the latter defines the melody. 
In ABC notation, each note is represented by a letter, and additional symbols are utilized to convey duration, rhythm, and other musical characteristics. 
An example is illustrated in Figure~\ref{fig:abc_sample}.
``V:1’’ indicates the beginning of the first music track and the lines before it are headers. 
A tune can consist of one or more tracks, each representing a distinct musical element within the composition.
The bars within each track are separated by bar line symbols like vertical lines (``$|$’’), which  refer to the standard bar line.

\begin{figure*}[ht]
    \centering
    \includegraphics[width=1\textwidth]{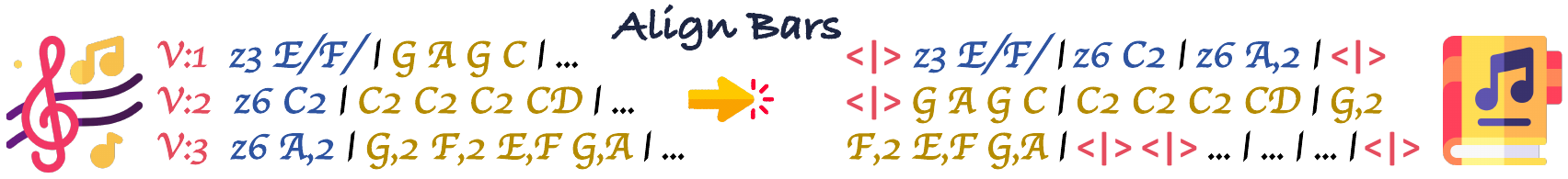}
    \caption{Illustration of synchronized multiple-track ABC notation. Music segments from bars sharing the same index across all tracks, along with their right bar lines, are concatenated to guarantee alignment. The combined elements are then enclosed by a pair of a newly introduced symbol ``\textless$|$\textgreater''.}
    \label{fig:CMMMU-sample}
\end{figure*}

In ~\citet{yuan2024chatmusician}, ABC files without any modification are the input of models. 
However, we found that the models struggle with bar alignment when dealing with multiple tracks. 
Since a track represents a section or division within a musical composition, such as one of the instrumental or vocal parts in a piece of music, it is crucial for models to capture the correspondence between tracks. 
Specifically, this correspondence exists in bars with the same indices, and thus, they should be treated as a series of groups.
To this end, we reorganize the tracks as depicted in Figure~\ref{fig:CMMMU-sample}. 
We concatenate music segments from bars with the same index across all tracks, including their right bar lines.
These concatenated elements from different tracks are then enclosed by a pair of a newly introduced symbol ``\textless$|$\textgreater'', which is not part of the original ABC system. 
This symbol represents the beginning or the end of a group of bars at the same stage. 
In cases where a tune contains only one track, each new unit will consist of a single bar. 
After processing all the bars, we obtain a synchronized version of the music notation, while the headers remain unchanged. 
The length of the tracks is not always identical due to repetition or other specific musical structures, which are difficult to handle exhaustively. 
Considering these special samples typically account for just a small portion (0.01\% in our dataset) of the entire dataset, we simply skip them in this scenario. 

\subsection{Tokenizer}
We chose YouTokenToMe (YTTM) \citep{youtokentome} framework to develop a tokenizer with a vocabulary of 50,000 tokens, leveraging the Byte-Pair Encoding (BPE) \citep{bpe} for ABC notation tokenization. This method is instrumental in segmenting the ABC text into manageable units, thereby enhancing the model's ability to interpret and process the input effectively. We do not apply any normalization and dummy prefix to the training corpus, without changing its form or adding extra parts at the beginning. 
Additionally, a unique symbol ``\textless$\text{n}$\textgreater``is employed to denote spaces within the ABC text, ensuring accurate space recognition by the model.
\begin{table}[h!]
\centering
\caption{MuPT model structure with different model size.}
\scriptsize
\label{tab:model_parameters}
\begin{tabular}{@{}lccccc@{}}
\toprule
Parameters               & 190M & 505M & 1.07B & 1.97B &  4.23B    \\ \midrule
Hidden size       & 768 & 1024 & 1280 & 1536 & 2048    \\
Layers                   & 12 & 16 & 20 & 24 &  32   \\
Feedforward hidden dimensions  & 3072 & 4096 & 5120 & 6144 & 8192 \\
Num heads                & 12 & 16 & 20 & 24 & 32   \\
Head size                & 256 & 256 & 256 & 256 & 256   \\ \bottomrule
\end{tabular}
\end{table}
\vspace{-5mm}

\subsection{Scaling Law}
\vspace{-5mm}
\begin{table}[!htp]\centering
\caption{Notation Definition for Scaling Law.}\label{tab:notation_def}
\scriptsize
\begin{tabular}{l|l}\toprule
\textbf{Notation} &\textbf{Definition} \\\midrule
$N$ &The number of parameters. \\
$D$ &The training tokens. \\
$U_{D}$ &The number of unique tokens used. i.e., Number of tokens in each epoch. \\
$A$, $ B$, $ E$, $ d$, $ k$, $ \alpha$, $ \beta$ &Term parameters requiring fitting in Scaling Laws. \\
$k_d$, $k_n$, $k_u$, $ k_{in}$ &Parameters to fit the term designed for overfitting after early stop points. \\
\bottomrule
\end{tabular}
\end{table}

The Chinchilla Law, proposed by DeepMind, is a scaling law that provides insights into the training of large language models (LLMs). Our experiments reveal that the Chinchilla Law \citep{hoffmann2022training} provides a good fit for general cases, where moderate models were trained with a moderate amount of data. In this section, we will list two improvements to Chinchilla Law for symbolic music scaling principles on limited training data.

\subsubsection{Optimizing Baseline Scaling Laws under Computational Constraints}
A pivotal aspect of scaling laws is the optimization of loss within the bounds of computational feasibility. This is formalized as minimizing the valid loss $L$, subject to constraints imposed by available computational resources ($C$), specifically FLOPs, as denoted below:

\begin{equation}
\arg\min_{N,D} L(N, D) \quad \text{s.t.} \quad \text{FLOPs}(N, D) = C
\end{equation}

This framework encapsulates the trade-offs between parameters ($N$) and training tokens ($D$), and decision-making processes inherent in scaling models under resource limitations, illuminating pathways to efficiency and efficacy in LLMs training. 
Notation definition is in Table~\ref{tab:notation_def}, and more details can be found in Appendix \ref{baseline-law}.


In this paper, we will use the Chinchilla Law\citep{hoffmann2022training} and Data-Constrained law\citep{muennighoff2024scaling} as baselines. The former is a classical baseline in LLMs' training and the latter is crucial to address the constraints faced in scenarios where the volume of available training data does not meet the ideal requisites. This phenomenon is typical in the music domain. Please refer to \ref{data-constrain-law} for more information. 

\subsubsection{Symbolic Music Scaling (SMS) Law}
\begin{figure}[h]
    \centering
    \includegraphics[width=0.75\textwidth]{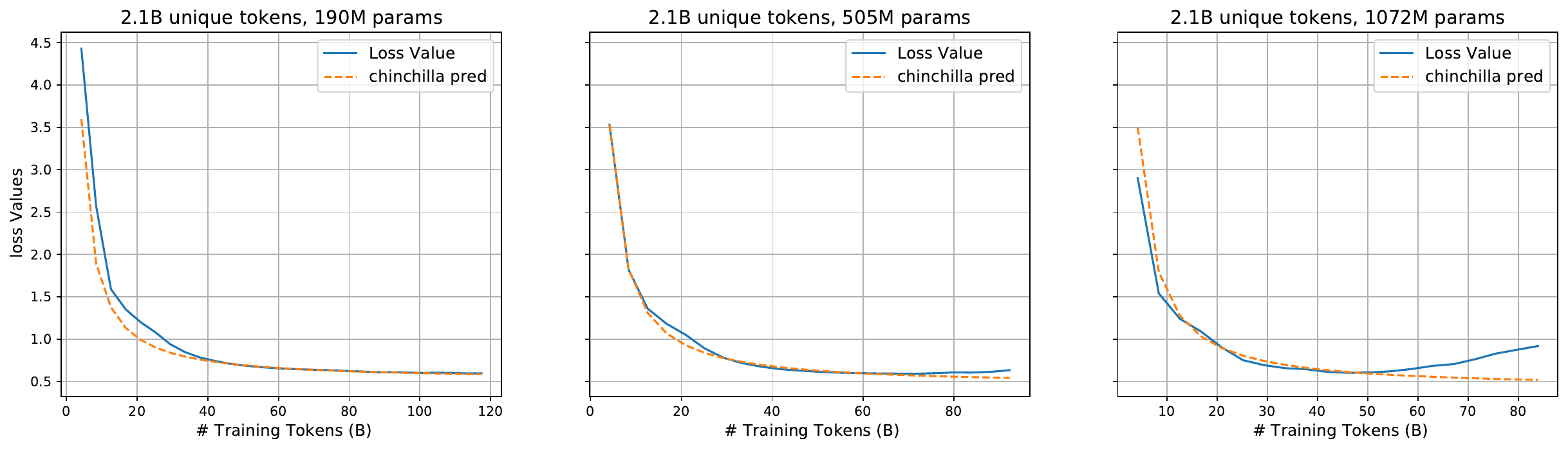}
    \caption{Chinchilla Law prediction and loss survey in the setting with 2.1B unique tokens.}
    \label{fig:mupt-chinchilla}
\end{figure}
Figure \ref{fig:mupt-chinchilla} demonstrates the Chinchilla prediction in yellow lines and the observed loss in blue. We can tell that the Chinchilla law does not provide good results when the data volume $D$ is small when the model just begins the pre-training stage, and when $D$ is large where repeated data provides overfitting. We proposed several modifications to address these problems.

\textbf{Continuous Adaptation of the Data-Constrained Law.}

The data volume $D'$ for Data-Constrained Law \cite{muennighoff2024scaling} at $n$ epoch is less then $D=n\times U_D$. We use its approximation of Equation 18 in Appendix A of their paper instead of the standard $D'$ for better prediction and simpler formulas. For more information please refer to Equation \ref{eq-D''} in Appendix \ref{D''}. We denote the modified data volume as $D''$.

\textbf{Incorporation of a New Term.}

We can observe that when that model parameter is small (e.g. $N=190M$), the Chinchilla underestimates the loss value and overestimates when the model size is large (e.g. $N=1072M$). This suggests that the coefficient $B$ in the Chinchilla formula $L = \frac{A}{N^\alpha} + \frac{B}{D^\beta} + E$ shall be relevant to $N$ instead of a constant. 

\begin{equation}\label{eq-1/ND}
L(N, D'') = \frac{d}{N^\alpha \cdot D''^\beta} + \frac{A}{N^\alpha} + \frac{B}{D''^\beta} + E.
\end{equation}

To address the model's limitations in accurately capturing performance metrics for smaller data sizes, we introduce an additional term, as delineated in Equation \ref{eq-1/ND}. This modification aims to refine the model's fidelity, particularly in scenarios characterized by limited data availability. Further details on this modification can be found in the Appendix \ref{1/ND}. After that, we proposed another term to predict the early stop points and overfited loss curve.

\textbf{Modelling Overfitting Settings.}

Crucially, previous iterations of the model fall short in predicting overfitting, particularly beyond early stopping thresholds. This gap is especially pronounced in the context of Data-Constrained environments, such as music, where open-source data is limited. To this end, we introduce a new component, $L_{overfit}$, to the model, encapsulated in Equation \ref{eq-MuPT}, to specifically account for overfitting losses:
\begin{equation}\label{eq-MuPT}
    L\left(N, D, U_D\right)=\frac{d}{N^\alpha\cdot D''^\beta} + \frac{A}{N^\alpha} + \frac{B}{D''^\beta} +E + L_{overfit}
\end{equation}
where $k_d$, $k_n$, $k_u$ and $k_{in}$ are constants, and 
\begin{equation} \label{eq-residule}
    L_{overfit} = GELU\left\{k_d\cdot D + k_n\cdot \log(N)-k_u\cdot \log(U_D) - k_{in}\right\}
\end{equation}

is our overfitting formulation. For comprehensive insights into the overfitting loss component, please refer to Appendix \ref{residule}.

\textbf{Parameter Fitting and Model Integration.}
 

Initial parameter fitting for $\{\alpha$, $\beta$, $A$, $B$, $E\}$, and $d$, subsequent linear regression analysis, focusing on the residuals between Equation \ref{eq-1/ND} and empirical observations, facilitates the calibration of overfitting parameters $\{k_d$, $k_n$, $k_u$, $k_{in}\}$ within Equation \ref{eq-residule}. The integration of these components in Equation \ref{eq-MuPT} not only predicts performance under constrained conditions but accounts for overfitting dynamics, helping to predict the true minimum of loss curve.

\section{Experiments}

\subsection{Experimental Setup}

As outlined in section \ref{sec:model arch}, we adopt similar model architecture from LLaMA2\citep{touvron2023llama2}, 
including RMSNorm\citep{zhang2019root} and SwiGLU\citep{shazeer2020glu}. 
In the full-scale data setting, we trained models of various sizes (ranging from 190M to 4.23B parameters) on the ABC text corpus, which consists of 33.6 billion tokens derived from a diverse collection of monophonic and polyphonic musical compositions spanning various genres and styles. 
For our data repetition experiments, we utilized subsets of the corpus, specifically 6.25\% and 25\% random sampled data.
The Adam\citep{kingma2014adam} optimizer and cosine learning rate schedules are applied throughout the training process.
All the hyperparameters are detailed in Appendix \ref{app:training details}.

\subsection{Scaling Law}
\subsubsection{Evaluation Metrics \& Fitting Methodology}
We use the $R^2$ value and Huber loss (with the parameter $\delta=1e-3$) between the authentic valid loss and predicted valid loss on small models (190M, 505M, 1.07B) to acquire the best scaling law. Then we use the best law to train two large models (with 1.97B and 4.23B). 
For more information about the two evaluation methods, please refer to Appendix \ref{evlauation}.


We optimized the SMS Law using the L-BFGS algorithm, the same with Chinchilla and Data-Constrained Laws. For more information, please refer to Appendix \ref{L-BFGS}.

\subsubsection{SMS Law are the Best on the Training Set}
\begin{table}[htbp]
\centering
\resizebox{0.9\columnwidth}{!}{
\begin{tabular}{c|cccc}
\toprule
\textbf{Paramatic fit} &  \textbf{$R^2$ Value (train) ↑} & \textbf{Huber Loss (train) ↓} &  \textbf{$R^2$ Value (test) ↑} & \textbf{Huber Loss (test) ↓} \\
\midrule
    Chinchilla law & 0.9347  & 0.0109 & -0.0933  & 0.0080\\
    Data-Constrained law & 0.7179 & 0.0206& 0.1524  & 0.0071\\
    Equation \ref{eq-D''} & 0.9075 & 0.0129 & 0.3114  & 0.0073\\
    Equation \ref{eq-1/ND} & 0.9759 & 0.0102 & 0.8580  & 0.0062\\
    SMS Law & \textbf{0.9780} & \textbf{0.0085} & \textbf{0.9612}  & \textbf{0.0028} \\
\bottomrule 
\end{tabular}
}
\caption{\label{table:Comparison}
Comparison of parametric fitting performance of different scaling laws.
}
\end{table}

The integration of an additional term as delineated in Equation \ref{eq-1/ND}, alongside the introduction of a GELU regularization component in Equation \ref{eq-residule}, collectively underpins the superior performance of the SMS Law, as empirically evidenced by its training set outcomes. This is particularly notable in the context of our parametric fitting performance comparison (see Table \ref{table:Comparison}), where the SMS Law outshines other scaling laws, achieving the highest $R^2$ value (0.9780) and the lowest Huber loss (0.0085) on the training set. 

Although Equation \ref{eq-D''} does not eclipse the Chinchilla Law in performance metrics, it nonetheless presents a significant improvement over the Data-Constrained Law's $D'$ by leveraging $D''$, which is indicative of a refined approach to managing the constraints posed by data repetition. This nuanced handling of data repetition, inherent to Equation \ref{eq-D''}, suggests an enhanced generalization capability in such scenarios. Therefore, we culminate it along with other modifications, manifest in the SMS Law in order to enhance model performance and generalization at the same time. In fact, it indeed provides much better results in the test set.

\subsubsection{Scaled-up Performance Followed SMS Law}
In our SMS Law experimentation under a computational budget of $1.2\times10^{21}$ FLOPs, we initially aim to train a 2.10B (or 1.98B) parameter model across 2.82 epochs on the whole 33.6B dataset per epoch, achieving a loss of 0.5279 (or 0.5280). Engineering constraints necessitated a slight scale-down to a 1.97 billion parameter model, which, intriguingly, showed a minimal loss increase to 0.529 around 2.5 epochs. 
Contrary to the predictions of SMS Law, the Chinchilla Law suggests optimal performance for a 990M parameter model over 6.1 epochs. Pushing boundaries, we continuously train the 1.07B parameter model and observe overfitting returns beyond 3 epochs, validating the SMS Law's advantages in this context.
Further, we train a 4.23B parameter model that underscored the SMS Law's predictive accuracy regarding overfitting risks, affirming its value as a strategic guide in scaling up models effectively within fixed computational constraints, beneficial for efficient model scaling decisions.


In validating the SMS Law, we analyze the performance of 1.97B and 4.23B parameter models, detailed on the right-hand side of Table \ref{table:Comparison}. This comparative study highlights the SMS Law's exceptional performance, evidenced by its unparalleled  $R^2$ values and minimal Huber Loss on testset as well. 

Unlike the Chinchilla and Data-Constrained laws, the SMS Law not only showcase superior predictive accuracy but also demonstrates its efficacy in optimizing neural network scaling within computational constraints. These results affirm the SMS Law's value in guiding scaling strategies for symbolic music, marking a significant advancement in the field.

\subsection{Evaluation}
\subsubsection{Efficiency of our training strategy}
\begin{figure*}[h!]
    \centering
    \caption{Training Loss for different model sizes and training strategy.}
    \includegraphics[width=1\textwidth]{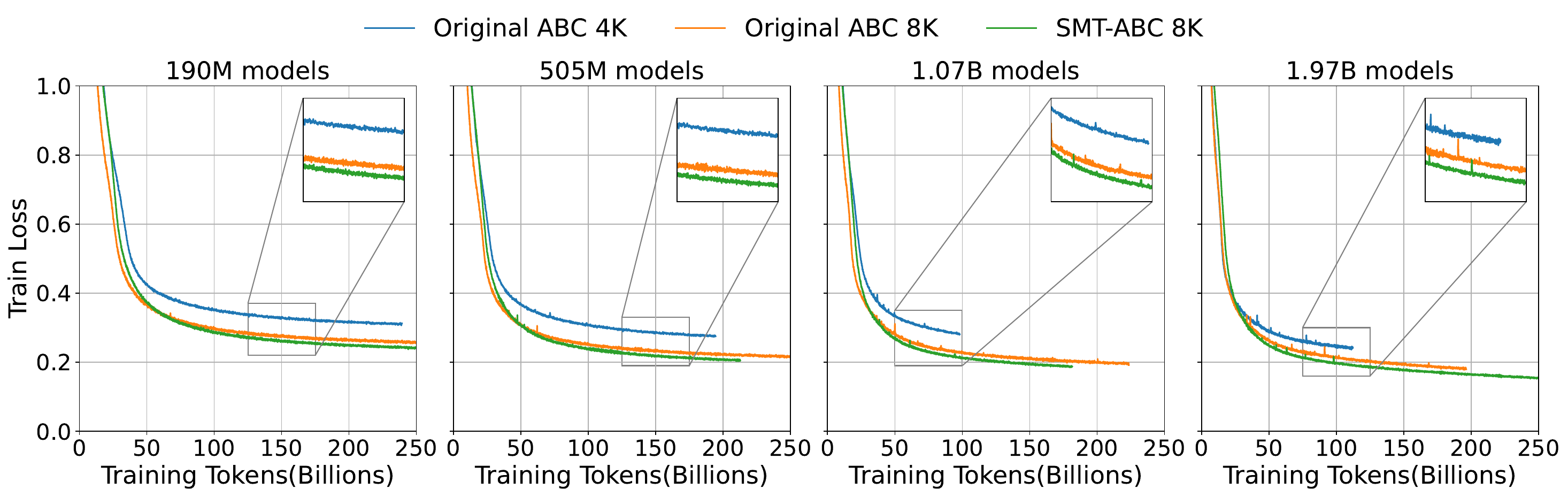}
    \label{fig:train_loss_curves}
\end{figure*}

To demonstrate the efficiency of our training strategies, we reference the training loss curves in Figure \ref{fig:train_loss_curves}. Our comparison spans four different model sizes: 190M, 505M, 1.1B, and 2B. We observed that increasing the training input length from 4096 to 8192 significantly reduces the loss, especially noticeable in the convergence phase. The comparison shows that after aligning data, our training loss slightly decreases compared to the original ABC loss, demonstrating our method's efficiency in improving training for various model sizes.

\subsubsection{Repetition Metrics}
\paragraph{Repetition rate} Repetition is significant in evaluating how well-structured the music is. In Table \ref{tab:combined_repet_similarity}, the piece-level average repetition rate of each system is calculated to reveal how often the repeat sign ${:|}$ appears in a generated set. It appears that 44.3\% of the generated samples from MuPT, which is quite close to the ground truth, and much higher than GPT-4. This suggests that MuPT is able to generate more music with repetition and structure.

\input{tables/combine_repet_similar}

\paragraph{Intra Similarity} In addition to the naive repetition rate, we also adopt the methods introduced in \citet{wang2024wholesong} to calculate the intra-similarity of music in each system. Specifically, a pre-trained VAE from \citet{yang2019disentanglement} and \citet{wang2020polyphonic} is transferred to compute the texture latent for each music piece; the intra-similarity of a music piece is defined as the average value of its texture latent similarity matrix, excluding the diagonal. Since the texture encoder is pre-trained on MIDI data, we transform ABC notations into MIDI format via the toolkit called abc2midi \footnote{https://github.com/xlvector/abcmidi} before the latent is obtained. Table \ref{tab:combined_repet_similarity} shows the mean value of each system's intra-similarity under the first-bar conditioned generation. MuPT achieves the highest score among all systems. Multitrack Music Transformers (\textit{MMT}) \citep{dong2023multitrack}, a MIDI-based music generation model, is also compared and its generated pieces have notably lower intra similarity than MuPT and GPT-4, both of which are ABC-based systems. This result corresponds with the intuition that score-level ABC notation is more capable of generating structured music than performance-level MIDI. 


\subsubsection{Subjective evaluation}

Human assessment should be involved to further testify the objective repetition metrics above. Following \citet{donahue2023singsong} and \citet{thickstun2023anticipatory}, we conduct a subjective listening study to measure the qualitative performance of MuPT against the ground truth (\textit{GT}) and baselines consisting of \textit{GPT-4}, \textit{MMT} and random note sequences (\textit{Random}). Listeners are asked to identify which of two musical excerpts from different sources is more "musical" during the test process. They are also instructed to focus on two aspects of musicality: how consistently the music sounds throughout (e.g., in terms of its melodic contours, rhythmic patterns, and chord progression); and how likely it is that the development of the music follows a clear structure (e.g., verse-chorus division, repetitions). This process is similar with that in \citet{yuan2024chatmusician} and its details are shown in the Appendix \ref{app:human_assess}.

\input{tables/combine_matrix_p-value}

Results for all systems are shown in Table~\ref{tab:combined_matrix_and_p-value}. Comparing our MuPT to GPT-4, listeners preferred music from our system in 79\% of cases. A Wilcoxon signed-rank test of these pairwise judgments indicates that listeners preferred music from MuPT significantly more often than MMT and GPT-4 ($p = 4.2249\times10^{-6}$ and $p = 6.6641\times10^{-8}$, respectively). 

\vspace{-0.3cm}
\section{Conclusion}
\vspace{-0.2cm}
In this paper, we introduce the MuPT series of pre-trained models for symbolic music generation, which set the standard of training open-source symbolic music foundation models. 
With 190M, 505M, 1.07B, 1.97B, and 4.23B parameters, these models have been pre-trained on the largest possible amount of ABC Notation data, including 33.6 Billion high-quality diverse symbolic music tokens.
Additionally, we dive deep into the scaling law exploration and propose SMS Law, a specialist in guiding future scaling of symbolic music foundation models.
Our results demonstrate that the MuPT series is competitive with mediocre human composers and guarantees state-of-the-art performance on symbolic music generation.
Moreover, MuPT introduces SMT-ABC, reordering the multiple-track original ABC notation format to assist pre-training of MuPT.
We believe that the open access of intermediate checkpoints of MuPT, SMS Law, and MuPT series will foster collaboration and innovation within the open-source computational music community, and open the door to the next-generation symbolic music foundation models.


\bibliography{colm2024_conference}
\bibliographystyle{colm2024_conference}

\newpage
\appendix

\section{Scaling Law}
\subsection{Scaling Law Baseline}\label{baseline-law}
\subsubsection{Abstracting Loss Metrics through the Chinchilla Law}

In this part, we focus on the relationship of loss metrics to various resource budgets in deep learning. It is first put forward by the Chinchilla Law as illustrated in Equation \ref{Chinchilla}. This law posits that both training and evaluation losses can be abstracted as a function of model capacity $N$ and training data size $D$, thus offering an insight to estimate the best combination of resources to be assigned to training.

\begin{equation}
L(N, D) = \frac{A}{N^\alpha} + \frac{B}{D^\beta} + E
\label{Chinchilla}
\end{equation}


Here, $L(N,D)$ denotes the loss metric during training or evaluation, which is assumed to exhibit a power-law dependency on $N$ and $D$. The parameters $A$, $B$, $E$, $\alpha$, and $\beta$ are determined by empirical fitting.

\subsubsection{Data-Constrained Law}\label{data-constrain-law}
\textbf{Data-Constrained Law: Scaling under Data Limitations.}
Complementing the Chinchilla Law, the Data-Constrained Law shows the scaling dynamics of LLMs when facing the data scarcity problem. Here, we strictly refer to the derivation method of \cite{muennighoff2024scaling}.
The goal of discovering Data-Constrained Scaling Law is to generalize the expression to multiple epochs where tokens are repeated. 

Data-constrained law is defined as:
\begin{equation} \label{data-constrain}
     L\left(N, D, U_D\right)=\frac{A}{N'^\alpha} + \frac{B}{D'^\beta} +E
\end{equation}
where
\begin{equation}\label{eq-D'}
\begin{split}
 & N' = U_N + U_NR_N^\star\left(1-\exp\left(\frac{-R_N}{R_N^\star}\right)\right)\\
 & D' = U_D + U_DR_D^\star\left(1-\exp\left(\frac{-R_D}{R_D^\star}\right)\right)\\
\end{split}
\end{equation}
To get a better understanding of the equation, the definitions of each of the above parameters are as follows:
Like Chinchilla Law, $ N $ is defined as the number of model parameters, and $ D $ is defined as the training tokens.

$U_D$ is defined as the number of unique tokens used. For data-constrained law, $ U_D $ is computed as min\{$D$,$D_C$\} given a budget of unique data $D_c$.

$U_N$ is defined as the number of ``unique'' parameters that provide an optimal fit for $U_D$. According to the method mentioned in \cite{muennighoff2024scaling}, given the following learned variables, $\{A,\alpha,B,\beta\,E\}$, the optimal allocation of compute(C) to $N$ and $D$ as follows:
\begin{equation}
    \begin{split}
 & N_{\text{opt}}(C) = G\left(\frac{C}{6}\right)^a \\
        & D_{\text{opt}}(C) = G^{-1}\left(\frac{C}{6}\right)^b \\
        & G = \left(\frac{\alpha A}{\beta B}\right)^{\frac{1}{\alpha+\beta}} \\
        & a = \frac{\beta}{\alpha+\beta} \\
        & b = \frac{\alpha}{\alpha+\beta}
    \end{split}
\end{equation}
Thus, $U_N$ is equal to $\min\{N_\text{opt},N\}$.

$ R_D $ is defined as the number of times the data is repeated. When training for a single epoch, $R_D=0$.

$R_N$ is the number that the `unique' parameters are repeated where $R_N = \max\{\left(\frac{N}{U_N}\right)-1,0\}$.

$D'$ is defined as the "effective data size": the number of unique data needed to get the same value as repeating $U$ unique tokens for $R_D$ repeats.The derivation process is as followed:

From a conceptual standpoint, the redundancy of data samples diminishes their incremental value in enhancing the model's knowledge base, given the model's prior exposure to said information. This principle underlies the hypothesis that each successive repetition of a sample contributes marginally less to the learning process, as the model has partially assimilated the information contained within the sample through prior iterations. To describe the process of training information loss, we have

\begin{equation}
\label{D'}
    D' = U + U\sum_{k=1}^{R_D}(1-\delta)^k\ = U+(1-\delta)U\frac{(1-(1-\delta))^{R_D}}{\delta}
\end{equation}

where $ \delta $ is defined as the `forgetting rate'. Each time a series of tokens is trained on a model, the model learns a $1-\delta$ fraction information from the optimization process. Assuming that the number of epochs beyond which repeating does not help, the right-hand side goes to to $\frac{(1-\delta)U}{\delta}$, since $\lim_{R_D\to\infty}(1-(1-\delta)^{R_D}) = 1 $. We define $R_D^\star $ is defined as$ \frac{1-\delta}{\delta}$, which is a learned constant.
According to Taylor expansion, if $\delta$ is small, we have:


\begin{equation}\label{eq-D'-new}\mathrm{e}^\frac{-1}{R_D^\star} \approx\ (1-\delta)\end{equation}

Now inserting$\frac{(1-\delta)}{\delta}=R^\star_D$ and $(1-\delta)^{R_D}=e^{(\frac{-1}{R_D^\star})^{R_D}}$ into Equation\ref{D'}, we get our final equation representing the effective data.

As the frequency of encountering repeated tokens diminishes, the benefit gained from processing them also decreases. Hence, the derivation of the $N'$ is similar to $D'$. In this context, there's no need to elaborate further. It should be pointed out that $R_N^\star$ is a learned parameter.

\subsection{Continuous Adaptation of the Data-Constrained Law.}\label{D''}
To enhance the predictive accuracy of the Data-Constrained law 
\citep{muennighoff2024scaling}
for continuous domains, we extend the original discrete formulation \ref{eq-D''} to accommodate continuous variables, allowing for a more nuanced understanding of data constraints in varied contexts. 
For an in-depth discussion on the derivation and implications of this continuous formulation, please refer to Appendix \ref{D''}.
\begin{equation}\label{eq-D''}
    L(N, D, U_D) = \frac{A}{N^\alpha} + \frac{B}{{D''}^\beta} + E
\end{equation}

where $k$ is a new parameter to be fit, and 
$D''$, the adjusted data size, is given by:
\begin{equation}
D''= \frac{1-k^{D/U_D}}{1-k}U_D.
\end{equation}

The definition of $D'$ in Equation \ref{D'} is defined from a discrete version and can not be extended to the case when D is less than $U_D$. So we reform the Equation \ref{D'} to
\begin{equation}
\begin{split}
    D'&=\frac{1-(1-\delta)^{\frac{D}{U_D}}}{\delta}\cdot U_D\\
    &=\frac{1-k_d^{\frac{D}{U_D}}}{1-k_d}\cdot U_D
\end{split}
\end{equation}
where $k_d:= 1-\delta$. This equation is equivalent to equation \ref{eq-D'-new} when $D$ is a positive integer times $U_D$.

We implemented a formula symmetric to $N'$ with $U_N$ and $k_N$. But the calculation results of $k_N\approx 0.999$. To make the formula simple, we use the original $N$ instead of $N'$ in the following formula.
\subsection{Motivation of SMS Law}
\subsubsection{Motivation of Adding Power of ``$ND$'' Term}\label{1/ND}
\begin{figure}[ht]
    \centering
    \includegraphics[width=0.5\textwidth]{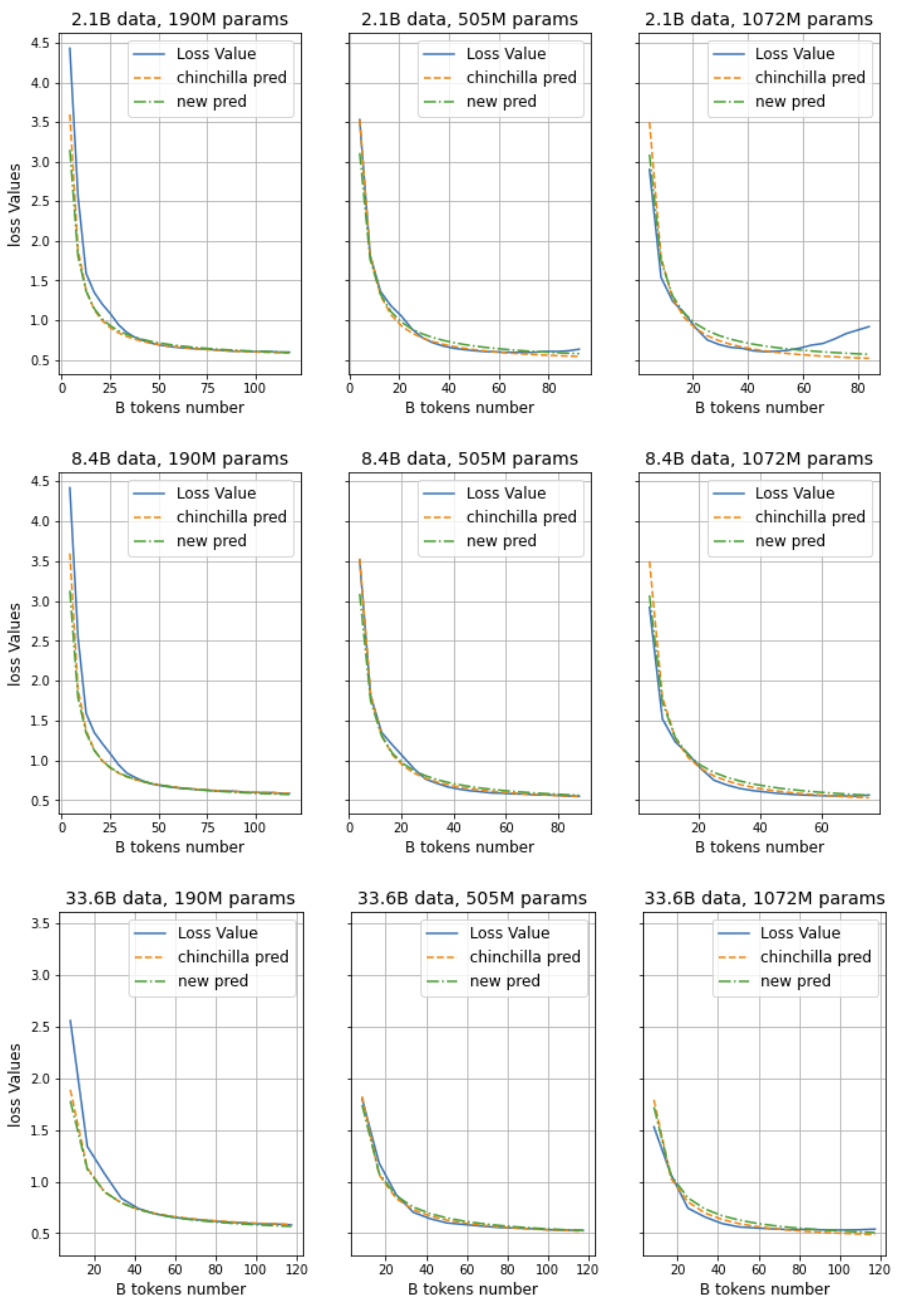}
    \caption{The loss curve, Chinchilla prediction, and Equation\ref{eq-D''} on 2.1B, 8.4B and 33.6B training data.}
    \label{fig:D''-curge}
\end{figure}
In our submission, we present an in-depth analysis of the model's loss dynamics as illustrated in Figure \ref{fig:D''-curge}, which juxtaposes the empirical loss trajectory (depicted through a blue line) against the theoretical predictions derived from the Chinchilla Law (illustrated by a yellow line) and further contextualized by Equation \ref{eq-D''}. This comparative study spans three distinct datasets—2.1B, 8.4B, and 33.6B data points—across models of varying capacities: 190M, 505M, and 1.07B parameters, respectively, arranged in a matrix of subfigures with datasets delineated by rows and model capacities by columns.

Observations across all data volumes reveal a nuanced interaction between model and data sizes. Specifically, for smaller datasets and model sizes (190M parameters), predictions consistently underestimate actual loss values, whereas for smaller datasets paired with larger models (1.07B parameters), predictions tend to overestimate. This discrepancy underscores a critical insight: loss reduction accelerates with increasing model size, suggesting a modified loss function, $\frac{A+\epsilon}{N^\alpha}$ over the simpler $\frac{A}{N^\alpha}$

Crucially, the term $\epsilon$ emerges as a function of a single variable $N$, ensuring variability in $\frac{\epsilon}{N^\alpha}$ across each unique model configuration shifting upwards or downwards without changing the shape. The ideal adjustment implies that $\epsilon$ approaches zero for large datasets, yet remains significant for smaller ones, highlighting its dependency on data volume $D$.

In addressing potential overfitting, our strategy focuses on minimizing parameter growth in line with Equation \ref{eq-D''}. A straightforward approach involves augmenting the loss $L$ into a polynomial encompassing $\frac{A}{N^\alpha}$ and $\frac{B}{D^\beta}$, with Equation \ref{eq-1/ND} introducing an additional term, $\frac{d}{N^\alpha\cdot D^\beta}$, to the existing framework. This refinement, while ostensibly simple, has been shown to yield robust and promising outcomes, exemplifying the efficacy of our proposed modifications in enhancing model performance within the context of scaling laws.

\subsubsection{Motivation of Linear Regression Term for Overfitted Residual}\label{residule}
\begin{figure}[ht]
    \centering
    \includegraphics[width=0.8\textwidth]{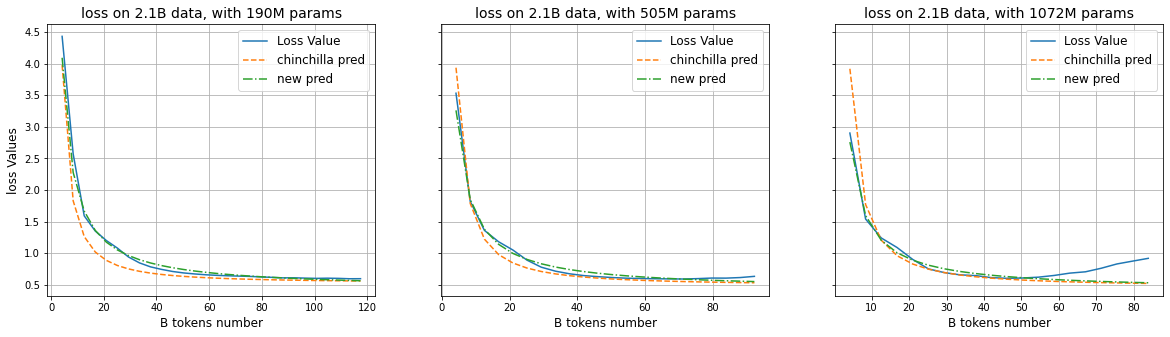}
    \caption{The loss curve, Chinchilla prediction, and Equation \ref{eq-1/ND} (green lines) on 2.1B training data.}
    \label{fig:overfit}
\end{figure}
Figure \ref{fig:overfit} offers a detailed exposition on the fidelity of Equation \ref{eq-1/ND} in capturing the loss trajectory across training sets of varied model capacities (190M, 505M, and 1.07B parameters). It is evident from the analysis that the equation adeptly mirrors the empirical loss curve across a broad spectrum of configurations, with the exception of scenarios characterized by concurrently large model sizes and token counts. A notable oversight in the literature is the scant consideration of loss dynamics beyond early stopping points, a consideration of paramount importance in music domain due to the inherent constraints on training data.

In addressing the challenges posed by modelling loss post-early stopping, our investigation delineates two distinct methodologies. The first approach involves the integration of a regularization term within $D''$, aimed at reducing its magnitude beyond the early stopping threshold. Despite its conceptual appeal, this method falls short of providing an adequate fit to the observed data. Alternatively, we explore the augmentation of the loss function $L$ with an additional term, engineered to be negligible when both $D$ and $N$ are minimal, yet incrementally assertive in influencing the loss trajectory after early stopping points. This latter strategy not only aligns more closely with empirical observations but also introduces a nuanced mechanism to accommodate the unique requirements of training in the music domain, thereby extending the utility and applicability of scaling laws within this context.

\begin{figure}[ht]
    \centering
    \includegraphics[width=0.8\textwidth]{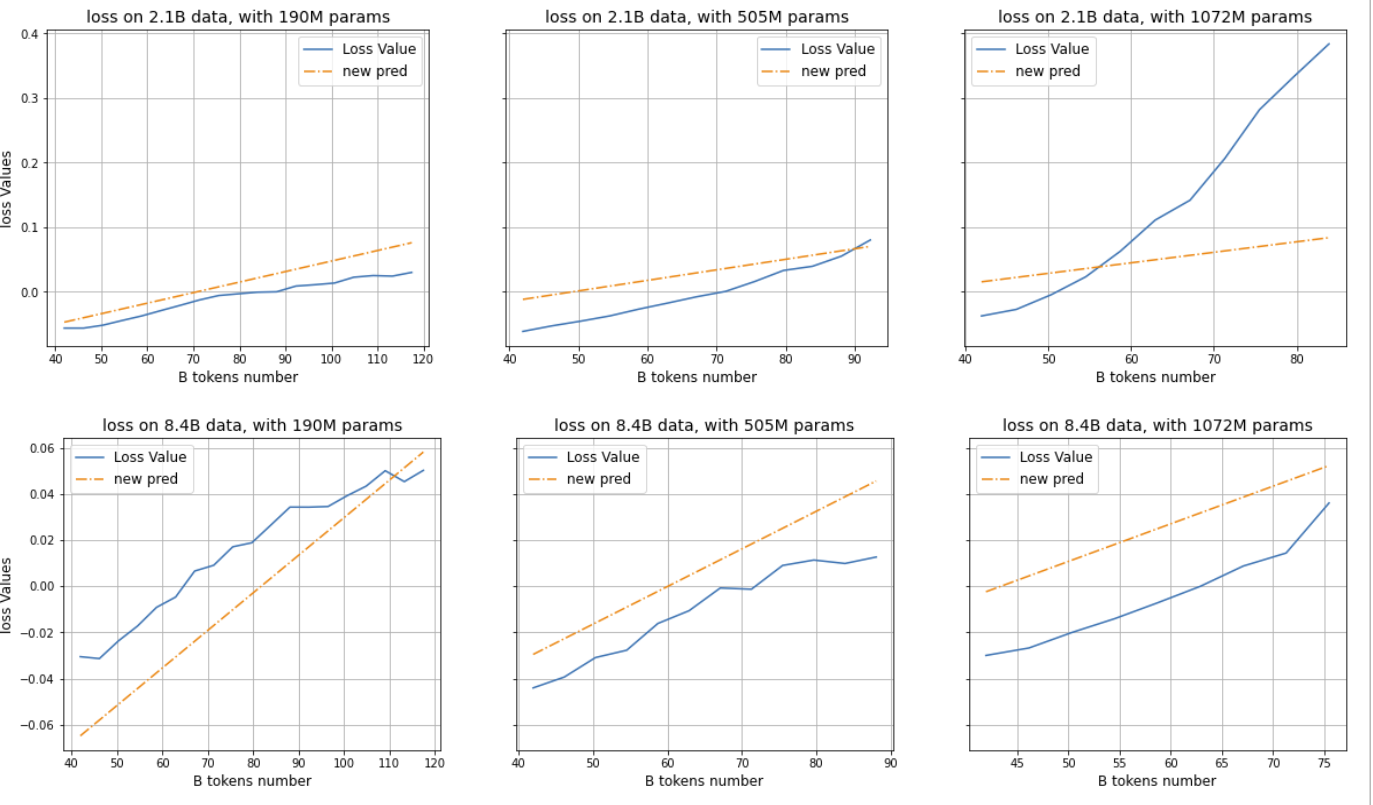}
    \caption{Residule between authentical valid loss and Equation \ref{eq-1/ND} prediction (blue lines), and the linear regression results (yellow lines).}
    \label{fig:residule}
\end{figure}
As delineated in Figure \ref{fig:residule}, the analysis of residuals post the 40 billion token threshold unveils a discernible onset of overfitting, which intriguingly appears to correlate with the model size, data capacity, and the count of unique tokens processed within a single epoch. This overfitting is further characterized by a linear dependency of loss on the total number of processed tokens, coupled with a quasi-linear transition of early stopping points observed across different model capacities (as organized in rows) and magnified across columns.

The progression of model capacities—doubling across rows and quadrupling across columns—illuminates a systematic pattern, suggesting that the early stopping points and consequently, the predicted loss, might be effectively modeled through a linear regression involving dataset size $D$, the logarithm of model capacity  $\log(N)$, and and the logarithm of unique tokens per epoch $\log(U_D)$. This observation culminates in the proposition of a regularization term formulated as $k_d\cdot D + k_n\cdot \log(N)-k_u\cdot \log(U_D) - k_{in}$, aimed at encapsulating and mitigating the observed overfitting dynamics.

\begin{table}[H]
    \centering
    \scalebox{1}{
    \begin{tabular}{lcc}
        \hline
        Activation Function  & $R^2$ (test)↑ & Huber Loss (test)↓\\
        \hline
        ReLU & \textbf{0.9786} & 0.0095 \\
        LeakyReLU & \textbf{0.9786} & 0.0095 \\
        GELU & 0.9780 & \textbf{0.0085} \\
        Tanh & \textbf{0.9786} & 0.0094 \\
        SELU &0.9779&0.010\\
        Sigmoid&0.6030&0.0700\\
        \hline
    \end{tabular}
    }
    \caption{Ablition study on the activation function.}
    \label{tab:Variables}
\end{table}

In addressing the intricacies of regularization within the context of early model training, especially when considering models of smaller scale (where $U_D$ and $D$ are minimal while $N$ is comparatively large), it becomes imperative to ensure that the regularization term does not adopt a substantially negative value. This stipulation aims to prevent undue penalization at the onset of training, thereby necessitating the incorporation of an activation function that tempers the regularization term's behavior. The Gaussian Error Linear Unit (GELU) function emerges as an apt choice in this scenario. GELU approximates the Rectified Linear Unit (ReLU) function for positive inputs, while also permitting slight negative values with minimal absolute magnitude, thus offering a balanced solution.

Empirical evidence, as detailed in our analysis, underscores the efficacy of applying the GELU function to the regularization term, notably achieving the lowest training loss alongside the second-highest $R^2$ value among the tested models. This finding is particularly salient given the broader magnitude of loss variations relative to $R^2$ values, thereby accentuating the GELU function's suitability for our regularization term. Consequently, the finalized model, incorporating the GELU-modulated regularization term, is depicted through a yellow line in Figure \ref{fig:residule}.
This strategic application of the GELU function not only mitigates the potential for excessive early training penalization but also optimizes the regularization term to enhance model performance effectively.

This approach not only elucidates the linear interdependencies among critical factors influencing model performance but also presents a nuanced regularization strategy designed to enhance model generalizability. Through the integration of this regularization term, we aim to establish a more robust and theoretically informed framework for predicting and managing loss trajectories in large-scale training regimes.
 
\subsection{Evaluation Metrics}\label{evlauation}
The R-squared value, also known as the "Coefficient of Determination," is a statistical measure used to evaluate the goodness-of-fit of a regression model. It is defined as:\begin{equation} R^2 = 1 - \frac{SS_{\text{res}}}{SS_{\text{tot}}} \end{equation}
Where $SS_{res}$ represents the Sum of Squares of Residuals, indicating the total sum of squared differences between the predicted values of the model and the actual observed values, $SS_{tot}$ represents the Total Sum of Squares, indicating the total sum of squared differences between the observed values of the dependent variable and their mean value.

The Huber loss is a type of loss function commonly employed in robust regression models. Unlike the squared error loss, which is sensitive to outliers in the data, the Huber loss is designed to be less affected by outliers. It achieves this by combining the characteristics of both the squared error loss and the absolute error loss. It is defined piecewise by:
\begin{equation}
Huber_\delta(y, f(x)) = \begin{cases} 
\frac{1}{2}(y - f(x))^2, & \text{if } |y - f(x)| \leq \delta \\
\delta(|y - f(x)| - \frac{1}{2}\delta), & \text{otherwise} 
\end{cases}
\end{equation}

For small residuals, it behaves like the squared error loss, whereas for large residuals, it behaves like the absolute error loss. This allows the Huber loss to provide a balance between the two, resulting in a more robust estimation procedure.

\subsection{Parameters Fitting Approach}\label{L-BFGS}
In our study, we adopt a methodology analogous to the Chinchilla Law and the Data-Constrained Law, employing the L-BFGS algorithm—a limited-memory quasi-Newton method—for the optimization of the Huber Loss. This loss function is applied between the logarithm of the predicted loss and the logarithm of the observed (authentic) loss across multiple runs. The objective is to identify the optimal parameters (best\_para) that minimize this Huber Loss, formalized as follows:
\begin{equation}
\begin{split}
    best\_para=\min\sum_{run i} Huber_\delta&\left\{\log\left[
        \frac{d}{N^\alpha\cdot D''^\beta} + \frac{A}{N^\alpha} + \frac{B}{D''^\beta} +E\right]_i, 
        \log(L_i)\right\}\\
    =\min\sum_{run i} Huber_\delta&\left\{
        LSE\left[\log\left(\frac{d}{N^\alpha\cdot D''^\beta}\right), \log\left(\frac{A}{N^\alpha}\right), \log\left(\frac{B}{D''^\beta}\right),\log(E)\right]_i, 
        \log(L_i) \right\}\\
    =\min\sum_{run i} Huber_\delta&\left\{
        LSE\left[\begin{split}
            &\log(d)-\alpha\log(N)-\beta\log(D'')\\
            &\log(A)-\alpha\log(N)\\
            &\log(B)-\beta\log(D'')\\
            &\log(E) 
        \end{split}\right], 
        \log(L_i) \right\}\\
\end{split}
\end{equation}
where $LSE$ refers to the \texttt{log-sum-exp} a numerically stable method to compute the logarithm of a sum of exponentials of inputs. The Huber Loss parameter, $\delta$ is set to $1e-3$, reflecting a stringent criterion for switching between squared loss and absolute loss to ensure robustness in optimization. Additionally, the L-BFGS algorithm's learning rate is configured at $1e-1$, 
with an update history size of 10 to balance between computational efficiency and the capacity to capture relevant optimization trends.

\subsection{Results of Proposed Methods with Early Stops}
\begin{table}[htbp]
\centering
\resizebox{0.8\columnwidth}{!}{
\begin{tabular}{c|cccc}
\toprule
\textbf{Paramatic fit} &  \textbf{$R^2$ Value (train) ↑} & \textbf{Huber Loss (train) ↓} &  \textbf{$R^2$ Value (test) ↑} & \textbf{Huber Loss (test) ↓} \\
\midrule
    Chinchilla law & 0.9443  & 0.0073 & -0.0004 & 0.0029\\
    Data-Constrained law & 0.7216 & 0.0189& 0.1005  & 0.0050\\
    Equation \ref{eq-D''} & 0.8356 & 0.0151 & 0.5829 & 0.0045\\
    Equation \ref{eq-1/ND} & 0.9843 & 0.0072 & \textbf{0.9866}  & \textbf{0.00088}\\
    SMS Law & \textbf{0.9851} & \textbf{0.0055} & 0.9864  & 0.00091 \\
\bottomrule 
\end{tabular}
}
\caption{\label{table:w-early-stop-Comparison}
Comparison parametric fitting performance of different Scaling Laws on the curve before early stop points.
}
\end{table}
From the table, we can see that most of the experimental results increase after we delete the curve after the early stop points. Adding the linear regression still contributes to the performance increase on the training set but provides worse results on test set compared to Equation \ref{eq-1/ND}.

\section{A Short Lecture Note of L-BFGS Algorithm}
BFGS (Limited-memory Broyden–Fletcher–Goldfarb–Shanno) is a variant of the BFGS method, a quasi-Newton optimization algorithm used to solve unconstrained nonlinear optimization problems. It is particularly suitable for handling large-scale optimization problems by limiting the size of the stored matrices, thus reducing storage and computational costs.

The core idea of the L-BFGS algorithm is to approximate the inverse of the Hessian matrix of the objective function using historical records of function values and gradients. In contrast to traditional Newton's method that requires storing and updating the complete Hessian matrix, L-BFGS method only needs to store and update some historical information, making it more efficient in terms of storage and computation. It iteratively constructs an approximate inverse Hessian matrix to update parameters and continuously optimize the objective function until reaching a local optimum or satisfying convergence criteria.

According to Newton-Raphson method:
\begin{equation} \label{data-constrain1}
\begin{split}
 & f:\mathbb{R}^n\to\mathbb{R} \\
 & f(x_t+d)=f(x_t)+\nabla f(x_t)^Td+\frac{1}{2}d^T\nabla^2f(x_t)d+o(\|d\|^2)\\
\end{split}
\end{equation}
\begin{equation}
 h(d):=f(x_t+d)=f(x_t)+\nabla f(x_t)^Td+\frac{1}{2}d^T\nabla^2f(x_t)d \\
\end{equation}

\begin{equation} \label{3}
\begin{split}
         & \hat{d}:=\arg\min_dh(d) \\
         & \nabla h(\hat{d}) = \nabla f(x_t)+\nabla f^2(x_t)\hat{d}=0 \\
\end{split}
\end{equation}

\begin{equation}
x_{t+1}=x_t+\hat{d}=x_t-\nabla^2f(x_t)^{-1}\nabla f(x_t)
\end{equation}
According to BFGS:
\begin{equation}
    B_{k+1}=B_k-\frac{B_ks_ks_k^TB_k}{s^T_kB_ks_k}+\frac{y_ky^T_k}{y_k^Ts_k}
\end{equation}

In the BFGS algorithm, storing the approximate Hessian matrix at each iteration can be costly in terms of memory, especially in high-dimensional data scenarios. However, in practical computation, what we primarily need is the search direction. To address this issue, the L-BFGS algorithm was introduced as an improvement over the BFGS algorithm.

In L-BFGS, instead of storing the full Hessian matrix, only the most recent iterations' information is retained, significantly reducing the memory footprint. 

let $\rho_k=\frac{1}{y^T_ks_k}$, $V_k=I-\frac{y_ks^T_k}{y_k^Ts_k}$, then $H_{k+1}$ can be represented as:
\begin{equation}
    H_{k+1}=V_k^TH_kV_k+\rho_ks_ks_k^T
\end{equation}
Note that $H_0=I$.
\begin{equation}
    \begin{split}
        & H_1=V_0^TH_0V_0+ \rho _0s_0s_0^T \\
        & H_2=V_1^TH_1V_1+ \rho _1s_1s_1^T \\
        &    =V_1^T(V_0^TH_0V_0+ \rho _0s_0s_0^T)V_1+ \rho _1s_1s_1^T \\
        &    =V_1V_0^TH_0V_0V_1+V_1^T \rho _0s_0s_0^TV_1+ \rho _1s_1s_1^T \\
        & \dots \\
        & H_{k+1}=(V_k^TV_{k-1}^T\cdots V_1^TV_0^T)H_0(V_0V_1\cdots V_{k-1}V_k) \\
        &    +(V_k^TV_{k-1}^T\cdots V_1^T)\rho_1s_1s_1^T(V_1\cdots V_{k-1}V_k) \\
        &    +\cdots \\
        &    +V_k^T \rho _{k-1}s_{k-1}s_{k-1}^TV_k \\
        &    + \rho _ks_ks_k^T \\
    \end{split}
\end{equation}
If only the first m steps are retained:
\begin{equation}
    \begin{split}
    & H_{k+1}=(V_k^T V_{k-1}^T \dots V_{k-m}^T) H_0 (V_{k-m} \dots V_{k-1} V_k) \\
    &    +(V_k^T V_{k-1}^T \dots V_{k-m}^T) \rho_1 s_1 s_1^T (V_{k-m} \dots V_{k-1} V_k) \\
    &    +\dots \\
    &    +V_k^T \rho_{k-1} s_{k-1} s_{k-1}^T V_k \\
    &    +\rho_k s_k s_k^T \\
    \end{split}
\end{equation}
Then only $s_k$ and $y_k$ is necessary to be remained.


\section{Training Details}\label{app:training details}

All the models are trained using Adam\cite{kingma2014adam}, with $\beta_1=0.9, \beta_2=0.95, eps=10^{-8}.$ We use a cosine learning rate schedule, decay the final learning rate from $3^{-5}$ to $3^{-6}$, with warmup ratio of 0.1. We apply a weight decay of 0.1 and gradient clipping of 1.0. Table \ref{tab:train_detail_appendix} shows other training details of each model. 

\begin{table}[h!]
\centering
\caption{Training Details for different ABC format and model settings.}
\label{tab:train_detail_appendix}
\resizebox{\textwidth}{!}{%
\begin{tabular}{cccccc}
\toprule
 & Parameters & Context Length & Trained Tokens &  Training Days & Num of GPUs  \\ 
 \midrule
\multirow{8}{*}{\textbf{Original ABC}} & 190M & 4096 & 119B & 8.4 & 2 \\
 & 505M & 4096 & 97B & 8.4 & 4 \\
  & 1.07B & 4096 & 49B & 8.3 & 4 \\
 & 1.97B & 4096 & 56B & 8.4 & 8 \\
 \cmidrule{2-6}
 & 190M & 8192 & 346B & 6.9 & 8 \\
 & 505M & 8192 & 322B & 4.1 & 32 \\
  & 1.07B & 8192 & 223B & 5.4 & 32 \\
 & 1.97B & 8192 & 196B & 8.1  & 32 \\
 \midrule
\multirow{4}{*}{\textbf{SMT-ABC}} & 190M & 8192 & 276B & 5.5 & 8 \\
 & 505M & 8192 & 212B & 2.7 & 32 \\
  & 1.07B & 8192 & 181B & 4.4 & 32 \\
 & 1.97B & 8192 & 272B & 11.3 &  32 \\
 & 4.23B & 8192 & 262B & 10.7  & 64 \\
 \bottomrule
\end{tabular}
}
\end{table}

\section{Human Assessment}\label{app:human_assess}
We use webMUSHRA toolkit \citep{schoeffler2018webmushra} to conduct a web-based subjective listening AB-test. During the listening test, we ask the participants to choose the better one between a pair of music excerpts generated from two randomly selected different systems from \textit{GT, MuPT, GPT-4, MMT and Random} by considering the "Musicality" which indicates the overall perceptive quality of the music. Participants are encouraged to make a choice by refering to the guidelines below:
\begin{itemize}
\item[$\bullet$] How consistent the music sounds as a whole (e.g., in terms of its melodic 
contours, rhythmic patterns, and chord progression).
\item[$\bullet$] How likely the development of the music follows a clear structure (e.g. 
verse-chorus division, repetitions).
\item[$\bullet$] If you cannot understand the two guidelines above, just choose the one from A and 
B you prefer.
\end{itemize}

\end{document}

%% file: tables/combine_repet_similar.tex

\begin{table}[H]
    \centering
    \scriptsize
    \scalebox{1}{
    \caption{Mean value of the intra-texture similarity and repetition rate of each system. ABC notation string generated by MuPT achieves higher intra-similarity than the ground truth as well as those generated by GPT-4.}\label{tab:combined_repet_similarity}
    \begin{tabular}{lcc}
        \toprule
        System & Texture similarity & Repetition Det. Rate (\%)\\
        \midrule
        MuPT & \textbf{0.4288} & \textbf{44.3}\\
        GT & 0.3729 & 43.5\\
        MMT & 0.1767 & - \\
        GPT-4 & 0.3614 & 16.9\\
        \bottomrule
    \end{tabular}
    }
    
\end{table}


%% file: tables/combine_matrix_p-value.tex
\begin{table}[t]
  \scriptsize
  \begin{minipage}{0.37\textwidth}
    \centering
    \includegraphics[width=\columnwidth]{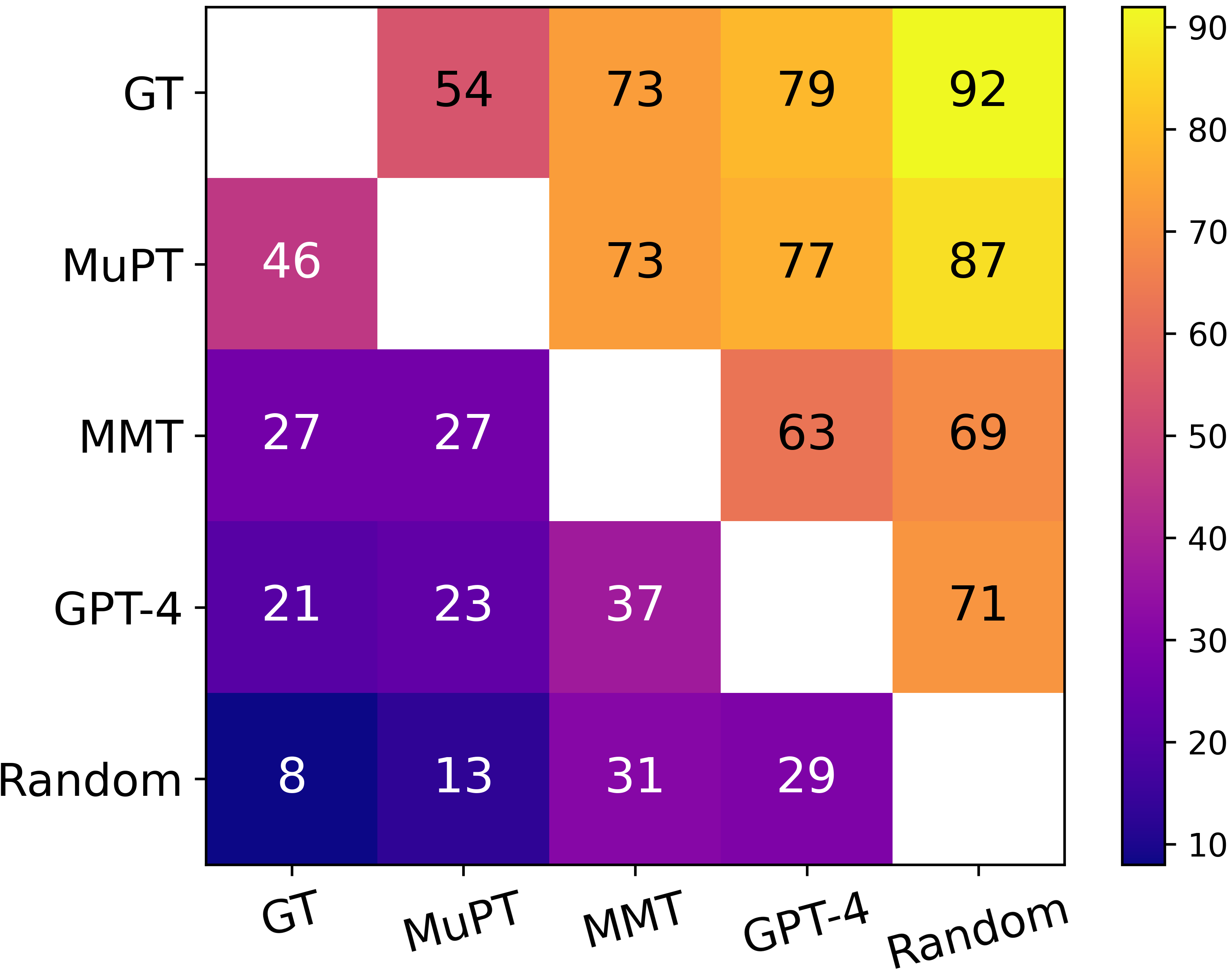}
  \end{minipage}%
  \begin{minipage}{0.65\textwidth}
    \centering
    \begin{tabular}{cccc}
        \toprule
        Model A & Model B & Wins (A/B) & p-value\\
        \midrule
        Human works & MuPT & 81/69 & 0.4237\\
        ~ & MMT & 109/41 & $4.2249 \times 10^{-6}$\\
        ~ & GPT-4 & 119/31 & $6.6315 \times 10^{-9}$\\
        ~ & Random & 138/12 & $4.4648 \times 10^{-17}$\\
        \midrule
        MuPT & MMT & 110/40 & $4.2249 \times 10^{-6}$\\
        ~ & GPT-4 & 115/35 & $6.6641 \times 10^{-8}$\\
        ~ & Random & 131/19 & $1.3618 \times 10^{-13}$\\
        \midrule
        MMT & GPT-4 & 95/55 & 0.0093\\
        ~ & Random & 103/47 & 0.0001\\
        \midrule
        GPT-4 & Random & 106/44 & $2.6691 \times 10^{-5}$\\
        \bottomrule
    \end{tabular}
  \end{minipage}
    \caption{Human evaluation of paired completions of musical excerpts generated by different sources given the first bar as the condition. The left is the matrix based on the AB test. Each row indicates the \% of times listeners preferred instrumentals from that system compared to those from each system individually (N = 150). Ground truth is denoted by GT. i.e.\texttt{77} means that listeners preferred MuPT over GPT-4 in 77\% of cases. The right is the absolute win numbers and the corresponding p-value of each pair. P-values are reported by a Wilcoxon signed rank test.}
  \label{tab:combined_matrix_and_p-value}
\end{table}
